\documentstyle[prl,aps]{revtex}
\begin{document}
\title{Realistic Neutrino Opacities for Supernova Simulations With Correlations and Weak Magnetism}
\author{C.~J.~Horowitz\footnote{e-mail:  charlie@iucf.indiana.edu} and
M.A. P\'{e}rez-Garc\'{\i }a\footnote{e-mail: angeles@niobe.iucf.indiana.edu}}
\address{Nuclear Theory Center and Dept. of Physics, Indiana
University, Bloomington, IN 47405}
\date{\today}
\maketitle
\begin{abstract}
Advances in neutrino transport allow realistic neutrino interactions to be incorporated into supernova simulations.  We add tensor couplings to relativistic RPA calculations of neutrino opacities.  Our results reproduce free-space neutrino-nucleon cross sections at low density, including weak magnetism and recoil corrections.  In addition, our opacities are thermodynamically consistent with relativistic mean field equations of state.  We find antineutrino mean free paths that are considerably larger then those for neutrinos.  This difference depends little on density.  In a supernova, this difference could lead to an average energy of $\bar\nu_\mu$ that is larger than that for $\nu_\mu$ by an amount that is comparable to the energy difference between $\nu_\mu$ and $\bar\nu_e$ 
\end{abstract}

\section{Introduction}

Core collapse supernovae are dominated by neutrinos, therefore many supernova properties may depend on the neutrino-nucleon interactions.  Recent advances in neutrino transport allow neutrino interactions to be accurately incorporated in simulations \cite{boltzmann,boltzmanna,boltzmannb}.   It is important to systematically improve the treatment of neutrino-nucleon interactions in order to take full advantage of the high quality neutrino transport and insure that modern simulations are as meaningful as possible.

There are fundamental differences between neutrino- and antineutrino-nucleon interactions related to parity and charge conjugation violation.  These differences were incorporated in refs. \cite{weakmag,weakmag2} by including weak magnetism corrections.  These corrections are important for both charged and neutral currents.  Weak magnetism has been included in supernova simulations with an average correction factor \cite{simulations,simulations2} and leads to a lower average energy for $\nu_\mu$ compared to $\bar\nu_\mu$ \cite{differences}.  The average neutrino energy for $\bar\nu_\mu$, $\bar\nu_\tau$ was found to be over one MeV larger then that for $\nu_\mu$, $\nu_\tau$ \cite{differences}.  This significant difference is comparable to the approximately 1.5 MeV difference in average energy of $\nu_x$ (all four $\bar\nu_\mu$, $\bar\nu_\tau$, $\nu_\mu$ and $\nu_\tau$ combined) compared to $\bar\nu_e$ found in a modern simulation that includes a variety of neutral current interactions such as nucleon bremsstrahlung and $\nu_e$, $\bar\nu_e$ annihilation \cite{nuxE}.  

Many-body corrections to neutrino interactions may also be important in the dense protoneutron star.  Nonrelativistic calculations \cite{nonrpa} find that correlations reduce the neutrino opacity at high densities, while relativistic calculations \cite{relrpa} find a similar but somewhat smaller reduction.   Unfortunately these corrections depend on the model employed for the strong interactions.  Furthermore, the model of the strong interactions used for the equation of state (EOS) needs to be consistent with the neutrino opacities to insure thermodynamic consistency.  For example, uniform neutron rich matter is unstable at low densities to a phase transition to a nonuniform crust phase.  This should show up as an enhancement of neutrino opacities as neutrinos excite low-energy density oscillation modes.

We believe the minimum requirements for accurate neutrino opacities are the correct low density limit and consistency with the EOS.  First, the neutrino interactions, at low densities, should agree with the exact free-space neutrino-nucleon cross sections including weak magnetism and recoil corrections.  This will insure that fundamental differences between neutrino and antineutrino interactions are included.  Second the opacities should be consistent with the EOS employed in the simulation.  This will insure that the opacities correctly reflect possible phase transitions in the EOS.  To our knowledge, these very minimum requirements have never been satisfied in a supernova simulation.

Most supernova simulations use the Lattimer Swesty EOS \cite{lseos} based on a nonrelativistic model for the strong interactions.  It is important to study the sensitivity of simulations to changes in the EOS.  In addition, we wish to include weak magnetism using a relativistic formalism.  Therefore we consider the EOS \cite{TM1eos} based on the TM1 relativistic mean field interaction \cite{TM1,TM1a}.  We calculate consistent neutrino opacities for this EOS in a relativistic RPA approximation.   Our starting point is the relativistic RPA calculations of Horowitz and Wehrberger at zero \cite{HW} and finite temperature \cite{HWT}.  Previous relativistic RPA calculations with the TM1 interaction did not include weak magnetism \cite{TM1RPA}.  Some additional relativistic calculations of neutrino opacities include \cite{otherrpa,othercharge,otherHF,otherHFa}. 

Section II presents our formalism for the neutrino cross section including weak magnetism corrections.  Results for mu and tau neutrino transport mean free paths are presented in section III as a function of density and temperature.  We also discuss differences between neutrinos and antineutrinos.  We focus on the low density region between the neutrino sphere and normal nuclear density.  These results are discussed and we conclude in Section IV.

\section{Formalism}

In this section we calculate neutrino scattering from neutron rich matter.  We assume only neutral current interactions which is appropriate for $\mu$ and $\tau$ neutrinos and antineutrinos.  We use a relativistic field-theoretical model for describing a system composed of protons, neutrons and electrons.  

In this work the TM1 model~\cite{TM1} is used where the nucleons interact via the exchange of $\sigma$,$\omega$  and $\rho$ mesons.  Non-linear scalar and vector self-interactions are included.  The TM1 nucleon-meson couplings are known to provide excellent descriptions of the ground states of heavy nuclei including unstable nuclei~\cite{TM1}.  The interaction Lagrangian, in our conventions is,
\begin{eqnarray}
{\cal L}_{\rm int} &=&
\bar\psi \left[g_{\rm s}\phi   \!-\! 
         \left(g_{\rm v}V_\mu  \!+\!
    \frac{g_{\rho}}{2}{\mbox{\boldmath $\tau$}}\cdot{\bf b}_{\mu} 
                               \!+\!    
    \frac{e}{2}(1\!+\!\tau_{3})A_{\mu}\right)\gamma^{\mu}
         \right]\psi \nonumber \\
                   &-& 
    \frac{\kappa}{3!} (g_{\rm s}\phi)^3 \!-\!
    \frac{\lambda}{4!}(g_{\rm s}\phi)^4 \!+\!
    \frac{\zeta}{4!}   g_{\rm v}^4(V_{\mu}V^\mu)^2 \;.
 \label{LDensity}
\end{eqnarray}
The model contains an isodoublet nucleon field ($\psi$) interacting
via the exchange of two isoscalar mesons, the scalar sigma ($\phi$)
and the vector omega ($V^{\mu}$), one isovector meson, the rho (${\bf
b}^{\mu}$), and the photon ($A^{\mu}$). The parameters are listed in Table I.

The differential cross section per nucleon for neutral current $\nu$-nucleon scattering is~\cite{HW},
\begin{eqnarray}
\frac {d^3\sigma}{d^2\Omega' dE'}  =
-\frac {G_F^2}{32\pi^2}
\frac{E'}{n_B E}~
\left[1-\exp{\left(-\frac{q_0}{T}\right)}\right]^{-1}~
{\rm Im}~(L^{\alpha \beta}\Pi^R_{\alpha \beta}) \,,
\label{cross}
\end{eqnarray}
where $n_B$ is the baryonic density and $E$ ($E'$) are the incoming (outgoing) neutrino energies.
The energy transfer is $q_0=E-E'$.  In terms of the incoming neutrino four momentum
$k_\mu =(E,\vec{k})$ and the four momentum transfer $q_\mu=(q_0,\vec{q})$, the lepton tensor $L_{\alpha\beta}$
is given by,
\begin{eqnarray}
L^{\alpha\beta}= 8[2k^{\alpha}k^{\beta}+(k\cdot
q)g^{\alpha\beta}
-(k^{\alpha}q^{\beta}+q^{\alpha}k^{\beta})\mp
i\epsilon^{\alpha\beta\mu\nu}
k_{\mu}q_{\nu}] \,
\end{eqnarray}
with a minus sign for neutrinos and a plus sign for antineutrinos.

\subsection{Hartree Response}

The retarded polarization tensor $\Pi^R_{\alpha\beta}$, is a function of the particles chemical potential $\mu_i$, temperature $T$, and the kinematical variables $q_0$, and $|\vec{q}|$.  For a plasma of protons $i=p$, neutrons $i=n$, and electrons $i=e$ it is,
\begin{eqnarray}
{\rm Im} \Pi^R_{\alpha\beta}
=\tanh{\left(\frac{q_0}{2T}\right)}\sum_{i=p,n,e} {\rm
Im}~\Pi^{(i)}_{\alpha\beta}  \,
\end{eqnarray}
where $\Pi^{(i)}_{\alpha\beta}$ is the time ordered or causal polarization and is given by,
\begin{eqnarray}
\Pi^{(i)}_{\alpha\beta}=-i \int
\frac{d^4p}{(2\pi)^4} {\rm Tr}~[T(G_i(p+q)\Gamma^i_{\alpha}
G_i(p)\Gamma^i_{\beta})]\, .
\end{eqnarray}
The Green's functions $G_i(p)$ describe the propagation of baryons ($i=p$ or $n$) or electrons ($i=e$) at finite
density and temperature in the several approximations considered.

\begin{table}
\caption{Parameters for the TM1 interaction using the conventions of Eq. (\ref{LDensity})}
\begin{tabular}{lccccccccc}
 $m_{\rm s} (MeV) $  & $g_{\rm s}^2$ & $m_{\rm v}$ (MeV) & $g_{\rm v}^2$ & 
           $\kappa$ (MeV) & $\lambda$ & $\zeta$ & $m_{\rho}$ (MeV)& $g_\rho$ \\
 511.198 & 100.57883 & 783 & 159.110473 & 2.829773 &  0.0003667 & 0.0169 &  770 & 85.8291 \\  
\label{Tableone}
\end{tabular}
\end{table}

In the Hartree or mean field approximation $G_i$ is \cite{HW,HWT},
\begin{equation}
G_i(p)=(\slash\hskip -5pt p_i^* + M_i^*)\Bigl\{ {1\over {p_i^*}^2 - {M_i^*}^2 + i\epsilon} + {i\pi\over E_{pi}^*} \delta(p_0^*-E_p^*) n_p^i\Bigr\},
\end{equation}
with $E_{pi}^*=({\bf p}^2 + {M_i^*}^2)^{1/2}$, $p_{0i}^*=p_0-V_i$ and ${\bf p}_i^*={\bf p}$.  The mean fields are,
\begin{eqnarray}
V_n &=& g_vV_0-g_\rho b_0/2, \\
V_p &=& g_vV_0+g_\rho b_0/2, \\
V_e &=& 0,
\end{eqnarray}
with $V_0$ the Omega (vector) and $b_0$ the Rho mean fields and $g_v$, $g_\rho$ the respective coupling constants.  The effective mass is $M_i^*=M^*=M-g_s\phi_0$ for $i=p,n$ and $M_e^*=M_e$. The scalar mean field is $\phi_0$ and the scalar coupling is $g_s$.  Finally $n_p^{(i)}$ gives the occupation of a state with momentum ${\bf p}$,
\begin{equation}
n_p^{(i)}=\Bigl\{1+{\rm exp}\bigl[(E_{pi}^*+V_i-\mu_i)/T\bigr] \Bigr\}^{-1}.
\end{equation}

The weak interaction vertex is $\Gamma^i_{\mu}$. It includes vector, axial, and tensor (or weak magnetism) contributions,
\begin{equation}
\Gamma^i_\mu(q)= F_1^i(Q^2)\gamma_\mu +
              iF_2^i(Q^2) \sigma_{\mu\nu} {q^\nu \over 2M} -
               G_A^i(Q^2)\gamma_\mu \gamma^5 \;. \quad
               (Q^2 \equiv {\bf q}^{2}-q_{0}^{2})\\
\end{equation}
The form factors $F_1^i, F_2^i, G_A^i$  are given in Table~\*{II}. For astrophysical purposes
we can neglect the $Q^2$ dependence since the medium is being probed at $Q^2\approx 0$.

In this paper we will be interested in the Hartree  (H) or Mean Field approximation
and in the Random Phase Approximation (RPA) where correlations are included.  The Hartree and RPA response functions which characterize the hot and dense medium, have been explicitly evaluated in previous works \cite{HW,HWT,TM1RPA} without considering the weak magnetism corrections.  We will discuss now the Hartree response functions. In this approach nucleon correlations are not taken into account.

The corrections introduced by the inclusion of the weak magnetism term will appear in the
polarization insertions and these can be separated into tensor, vector-tensor, and tensor-axial
pieces which must be calculated in addition to the other well known vector, and axial-vector parts extensively quoted in previous works\cite{HW,HWT}.
The polarizations induced by the weak magnetism in the Hartree approximation are:
\begin{eqnarray}
\Pi_{tt} ^{(i)\mu\nu}&=&-i \int{d^4p\over (2\pi)^4}\,
{\rm Tr}[G_i(p+q)\,{i\sigma^{\mu\alpha}q_\alpha\over 2M} \,G_i(p)\,
{-i\sigma^{\nu\beta}q_\beta\over 2M}]\label{poltt}\ \\
\Pi_{vt} ^{(i)\mu\nu} &=& -i \int{d^4p\over (2\pi)^4}\,
{\rm Tr}[G_i(p+q)\,\gamma^\mu \,G_i(p)\,
{-i\sigma^{\nu\beta}q_\beta\over 2M}]\label{polvt}\ \\
\Pi_{at} ^{(i)\mu\nu} &=& -i \int{d^4p\over (2\pi)^4}\,
{\rm Tr}[G_i(p+q)\,(-\gamma^\mu \gamma^5)\,G_i(p)\,
{-i\sigma^{\nu\beta}q_\beta\over 2M}]\label{polat}\
\end{eqnarray}
These are similar to the original vector-vector polarization \cite{HW},
\begin{equation}
\Pi_{vv}^{(i)\mu\nu} = -i \int{d^4p\over (2\pi)^4}\,
{\rm Tr}[G_i(p+q)\, \gamma^\mu\,G_i(p)\,\gamma^\nu ]\label{polvv}\,.
\end{equation}
More information on these polarizations is in the appendix.  At the Hartree level, the only modifications from the medium are Pauli blocking, and the density dependence of the nucleon effective mass.

\begin{table}
\caption{ Coupling constants with $g_a=1.260$, $\sin^2\theta_w=$
0.231, $\mu_p=$ 1.793 and $\mu_n=-$ 1.913.}
\begin{center}
\begin{tabular}{|c|c|c|c|}
\hline
 $Reaction$ & $F_1$ & $G_A$ & $F_2$ \\
\hline\hline
$\nu p\rightarrow \nu p$ & $ {1 \over 2} -2 \sin^2 \theta_{w}$  & $g_a /2$  &
${1\over 2} ( \mu_p -\mu_n ) - 2 \sin^{2} \theta_{w} \mu_p$  \\
$\nu n \rightarrow \nu n$ & $ -{1 \over 2} $  & $-g_a /2$  &
$ -{1\over 2} ( \mu_p -\mu_n ) - 2 \sin^{2} \theta_{w} \mu_n $ \\
$\nu e\rightarrow \nu e$ & $ 2 \sin^2 \theta_{w} -{1 \over 2}$ & $ -{1 \over 2}$ & $0$ \\
\hline
\end{tabular}
\end{center}
\label{table1}
\end{table}

The cross section can be written as,
\begin{equation}
{ d^3\sigma\over d^2\Omega' dE' } = { G_F^2 E' q_\mu^2 \over 4\pi^3 E n_B } 
\Bigl[ 1-{\rm exp}(-{q_0\over T})\Bigr]^{-1} 
\Bigl( A R_1 + R_2 \pm B R_3 \Bigr) ,
\label{sighar}
\end{equation}
with the plus sign for neutrinos, the minus sign for antineutrinos, and the kinematic factors,
\begin{equation}
A={2E E' +{1\over 2} q_\mu^2\over {\bf q}^2 },
\end{equation}
and,
\begin{equation}
B=E+E'\, .
\end{equation}
The three response functions are,
\begin{equation}
R_1 = \sum_{i=p,n,e}\bigl\{({F_1^i}^2 + {G_A^i}^2 )\ [{\rm Im}\, \Pi^{(i)}_{vvL}+{\rm Im}\, \Pi^{(i)}_{vvT}] +
2 F_1^i F_2^i\ [{\rm Im}\,\Pi^{(i)}_{vtL}+{\rm Im}\,\Pi^{(i)}_{vtT}]+{F_2^i}^2\ [{\rm Im}\,\Pi^{(i)}_{ttL}+{\rm Im}\,\Pi^{(i)}_{ttT}]\bigr\}\,,
\label{r1}
\end{equation}
\begin{equation}
R_2 = \sum_{i=p,n,e}[({F_1^i}^2 + {G_A^i}^2 )\ {\rm Im}\,\Pi^{(i)}_{vvT} +
2 F_1^i F_2^i\ {\rm Im}\,\Pi^{(i)}_{vtT}+{F_2^i}^2\ {\rm Im}\,\Pi^{(i)}_{ttT}-{G_A^i}^2 {\rm Im}\,\Pi_A^{(i)}]\,,
\label{r2}
\end{equation}
\begin{equation}
R_3 = \sum_{i=p,n,e}2(F_1^i+F_2^i {M^*_i \over M_i})\ G_A^i\ |{\bf q}|\ {\rm Im}\,\Pi^{(i)}_{va}\,.
\label{r3}
\end{equation}
Hartree cross sections based on Eq. (\ref{sighar}) with Eqs. (\ref{r1},\ref{r2},\ref{r3}) will be used to calculate neutrino mean free paths in Section III.

\subsection{RPA response}
We now discuss RPA corrections to the Hartree response functions from nucleon correlations.
In a mean field approximation for the ground state of neutron rich matter, the meson fields are calculated self-consistently.  This can be done with a Dyson equation that sums the tadpole diagrams.  The linear response of this system can be obtained by attaching the weak interaction to all nucleon lines.  This yields the RPA equations that sum the ring diagrams to calculate the nuclear response.  Calculating cross sections in RPA yields neutrino opacities that are thermodynamically consistent with an equation of state calculated in a mean field approximation, see for example \cite{TM1RPA}.  Therefore, the RPA opacities from this section are appropriate for supernova simulations involving the relativistic mean field TM1 equation of state \cite{TM1eos}.

In the RPA, correction terms from correlations are added to the Hartree responses of Eqs. (\ref{r1},\ref{r2},\ref{r3}).
\begin{equation}
R_1^{RPA} = R_1 + \Delta R_L + \Delta R_T + \Delta R_{aa}
\end{equation}
\begin{equation}
R_2^{RPA} = R_2 + \Delta R_T + \Delta R_{aa}
\end{equation}
\begin{equation}
R_3^{RPA} = R_3 + \Delta R_{va}
\end{equation}
The cross section in RPA is,
\begin{equation}
{ d^3\sigma\over d^2\Omega' dE' } = { G_F^2 E' q_\mu^2 \over 4\pi^3 E n_B } 
\Bigl[ 1-{\rm exp}(-{q_0\over T})\Bigr]^{-1} 
\Bigl( A R_1^{RPA} + R_2^{RPA} \pm B R_3^{RPA} \Bigr)\,.
\end{equation}

The correction $\Delta R_T$ from transversely polarized mesons and photons was discussed in ref. \cite{HW}.  Here we add weak magnetism corrections, the $\rho$ meson, and nonlinear meson self-couplings.
\begin{equation}
\Delta R_T = {\rm Im}\Bigl\{ \Pi^t_{TW} ({\bf 1} - {\bf D}_T {\bf \Pi}_T )^{-1} {\bf D}_T \Pi_{TW} \Bigr\}
\end{equation}
The polarization ${\bf \Pi}_T$ is a $3\times 3$ matrix for the system of electrons, protons and neutrons,
\begin{equation}
{\bf \Pi}_T= \left[\begin{array}{ccc} \Pi_{vvT}^e & 0 & 0
\\
0 & \Pi_{vvT}^p & 0
\\
0 & 0 & \Pi_{vvT}^n 
\end{array}\right],
\end{equation}
and the interaction matrix is,
\begin{equation}
{\bf D}_T = \left[\begin{array}{ccc} -\chi_\gamma & \chi_\gamma & 0 
\\
\chi_\gamma & -\chi_\gamma - \chi_v - \chi_\rho & \chi_\rho - \chi_v
\\
0 & \chi_\rho - \chi_v & -\chi_v - \chi_\rho
\end{array}\right]\,.
\end{equation}
The photon and meson propagators are,
\begin{equation}
\chi_\gamma = -{e^2\over q_\mu^2},
\end{equation}
\begin{equation}
\chi_v = -{g_v^2\over q_\mu^2 - {m_v^*}^2},
\end{equation}
\begin{equation}
\chi_\rho = -{{1\over 4}g_\rho^2\over q_\mu^2 - m_\rho^2},
\end{equation}
where because of self-interactions the vector meson effective mass is,
\begin{equation}
{m_v^*}^2 = m_v^2 + {\zeta\over 2}g_v^2 (g_v V_0)^2\,.
\end{equation}
Here $V_0$ is the vector meson mean field and the coupling constants are $e^2(e^2/4\pi=\alpha)$, $g_v^2$ and $g_\rho^2$.

The coupling of the system to weak currents is described by a three component vector $\Pi_{TW}$,
\begin{equation}
\Pi_{TW}= \left[\begin{array}{c}
F_1^e \Pi_{vvT}^e
\\
F_1^p \Pi_{vvT}^p + F_2^p \Pi_{vtT}^p
\\
F_1^n \Pi_{vvT}^n + F_2^n \Pi_{vtT}^n
\end{array}
\right],
\end{equation}
and its transpose $\Pi_{TW}^t$.  This vector contains the weak magnetism corrections involving $F_2^i$.

The correction from axial currents $\Delta R_{aa}$ is given by Eq. (50) of Ref. \cite{HW} with no corrections from weak magnetism.  
\begin{equation}
\Delta R_{aa} = {\bf q}^2 {\rm Im} \Bigl\{ \Pi_{aW}^t ({\bf 1} - {\bf D}_T {\bf \Pi}_T)^{-1} {\bf D}_T \Pi_{aW} \Bigr\}
\end{equation}
Here $\Pi_{aW}$ is a three component vector describing the coupling to weak axial currents and $\Pi_{aW}^t$ is its transpose,
\begin{equation}
\Pi_{aW}= \left[\begin{array}{c}
G_A^e \Pi_{va}^e
\\
G_A^p \Pi_{va}^p 
\\
G_A^n \Pi_{va}^n 
\end{array}
\right].
\end{equation}
Note, we assume no additional corrections to $\Delta R_{aa}$ from pion exchange or short range correlations that are often parameterized with a constant $g'$.  These can significantly reduce the RPA cross sections at high density \cite{relrpa}.  Therefore our results without $g'$ represent a minimal calculation that is consistent with the TM1 equation of state.  Clearly additional RPA interactions (such as $g'$), that do not contribute to the mean field equation of state, can be added without destroying the consistency. We discuss this further in section IV.

The vector-axial correction $\Delta R_{va}$ is,
\begin{equation}
\Delta R_{va}=2{\rm Im} \Bigl\{ \Pi_{aW}^t ({\bf 1} - {\bf D}_T {\bf \Pi}_T )^{-1} {\bf D}_T \Pi_{TW} \Bigr\}\,.
\end{equation}  

Finally $\Delta R_L$ describes corrections from longitudinally polarized mesons and photons.  This correction is slightly more complicated because of mixing in the medium between scalar mesons and longitudinally polarized vector mesons.  This can be taken into account by increasing the dimension of the matrices from $3\times 3$ to $4\times 4$.  The four rows describe vector coupling to electrons, protons and neutrons and a single scalar coupling.  Because the weak interaction has no scalar coupling one can combine all of the scalar polarizations into a single entry that is the sum of proton and neutron contributions.  This yields \cite{HW},
\begin{equation}
\Delta R_L = {\rm Im} \Bigl\{ \Pi_{LW}^t ({\bf 1}- {\bf D}_L {\bf \Pi}_L )^{-1} {\bf D}_L \Pi_{LW} \Bigr\}\,.
\end{equation}
Here the longitudinal polarization matrix is,
\begin{equation}
{\bf \Pi}_L= \left[\begin{array}{cccc} \Pi_{vvL}^e & 0 & 0 & 0
\\
0 & \Pi_S & \Pi_M^p & \Pi_M^n 
\\
0 & \Pi_M^p & \Pi_{vvL}^p & 0
\\
0 & \Pi_M^n& 0 & \Pi_{vvL}^n 
\end{array}\right].
\end{equation}
The scalar polarization $\Pi_S$ is the sum of proton and neutron contributions $\Pi_S=\Pi_S^p+\Pi_S^n$ with,
\begin{equation}
\Pi_S^{(i)}= -i\int {d^4p\over (2\pi)^4} {\rm Tr} [G_i(p+q) G_i(p)],
\end{equation}
and the mixed scalar-vector polarization is,
\begin{equation}
\Pi_M^{(i)}= -i \bigl({Q^2\over {\bf q}^2}\bigr)^{1/2} \int {d^4p\over (2\pi)^4} 
{\rm Tr} [G(p+q)_i\gamma_0 G_i(p)].
\end{equation}
The interaction matrix is,
\begin{equation}
{\bf D}_L = \left[\begin{array}{cccc} \chi_\gamma & 0 & -\chi_\gamma & 0 
\\
0 & -\chi_s & 0 & 0
\\
-\chi_\gamma & 0 & \chi_\gamma + \chi_v + \chi_\rho & \chi_v-\chi_\rho 
\\
0 & 0 & \chi_v-\chi_\rho  & \chi_v + \chi_\rho
\end{array}\right],
\end{equation}
with the scalar meson propagator,
\begin{equation}
\chi_s = -{g_s^2\over q_\mu^2 - {m_s^*}^2 }\,.
\end{equation}
The scalar effective mass is,
\begin{equation}
{m_s^*}^2 = m_s^2 + g_s^2[\kappa g_s\phi_0 +{\lambda\over 2} (g_s\phi_0)^2]\,.
\end{equation}
Finally, the longitudinal coupling of the system to the weak current is a four component vector,
\begin{equation}
\Pi_{LW}= \left[\begin{array}{c}
F_1^e \Pi_{vvL}^e
\\
(F_1^p + F_2^p{q_\mu^2\over 4MM^*})\Pi_M^p + (F_1^n + F_2^n{q_\mu^2\over 4MM^*})\Pi_M^n
\\
F_1^p \Pi_{vvL}^p + F_2^p \Pi_{vtL}^p
\\
F_1^n \Pi_{vvL}^n + F_2^n \Pi_{vtL}^n
\end{array}
\right],
\label{mixedW}
\end{equation}
and its transpose is $\Pi_{LW}^t$.  Note, the second line uses results for the mixed scalar-tensor polarization, see the appendix.  It is a simple matter to numerically evaluate the corrections $\Delta R_L$, $\Delta R_T$, $\Delta R_{aa}$ and $\Delta R_{va}$ to calculate the RPA cross section.

\subsection {$\nu$ Opacities}
We now calculate the transport mean free path for $\mu$ or $\tau$ neutrinos or antineutrinos in dense neutron rich matter. We assume the opacity is dominated by neutral current scattering from protons, neutrons and electrons. Well inside the neutrino sphere, the mean free path becomes much smaller than the size of the system, and we expect a simple diffusion approximation to be valid.  Using the diffusion approximation, the energy flux ${\bf F}$ can be written
\begin{equation}
{\bf F}=-\frac{<\lambda>}{3}{\bf \nabla} U
\end{equation}
where U is the energy density of the neutrinos.

The transport cross section for a neutrino of energy $E$ scattering through an angle $\theta$ is,
\begin{equation}
\sigma^t(E)=\int_0^\infty dE' \int d\Omega' \frac {d^3\sigma}{d^2\Omega' dE'} (1-cos(\theta)), 
\end{equation}
and the mean free path is simply,
\begin{equation}
\lambda(E) = \frac {1}{ \sigma^t(E) n_b}.
\end{equation}
Note, when evaluating the integration over the outgoing neutrino energy $E'$, the energy transfer $E-E'$ takes both positive and negative values.

The mean free path averaged over the incoming neutrino energy spectrum is,
\begin{eqnarray}
<\lambda>=\frac{\int_{0}^{\infty} \lambda(E)E^4f(E)dE}{\int_{0}^{\infty} E^4f(E)dE},
\label{eave}
\end{eqnarray}
For simplicity we assume $f(E)\approx {\rm e}^{-E/T}$ at temperature $T$.  The mean free path for a neutrino of average energy, $\lambda(<E>)$ with  $<E>=\pi T$, and the mean free path averaged over energy $<\lambda>$ are in general somewhat different. 

To quantify the differences between antineutrinos and neutrinos, we define a ratio,
\begin{eqnarray}
R=\frac{<\overline{\lambda}>-<\lambda>}{<\lambda>}
\label{R}
\end{eqnarray}
with $<\overline{\lambda}>$ the antineutrino and $<\lambda>$ the neutrino mean free path.  This difference comes mostly from weak magnetism corrections.

\section{Results}

We now present results for mu and tau neutrinos and antineutrinos scattering in neutron rich matter.  At low densities and temperatures there is a transition from a uniform liquid to a nonuniform solid phase.  This solid phase is expected to make up the crust of a neutron star.  The instability of the uniform system to density oscillations shows up as an enhancement in the RPA cross section for small values of the energy transfer $q_0=E-E'$ where $E$ is the initial and $E'$ the final neutrino energy.  Note, all of our calculations assume a uniform system.  Therefore they are not directly valid in the nonuniform phase.  However, the enhancement of the cross section we find in an RPA approximation could be a first hint of the perhaps greatly enhanced neutrino opacities we expect for the nonuniform phase, see below.

\begin{figure}
\vbox to 3.9in{\vss\hbox to 8in{\hss {\includegraphics{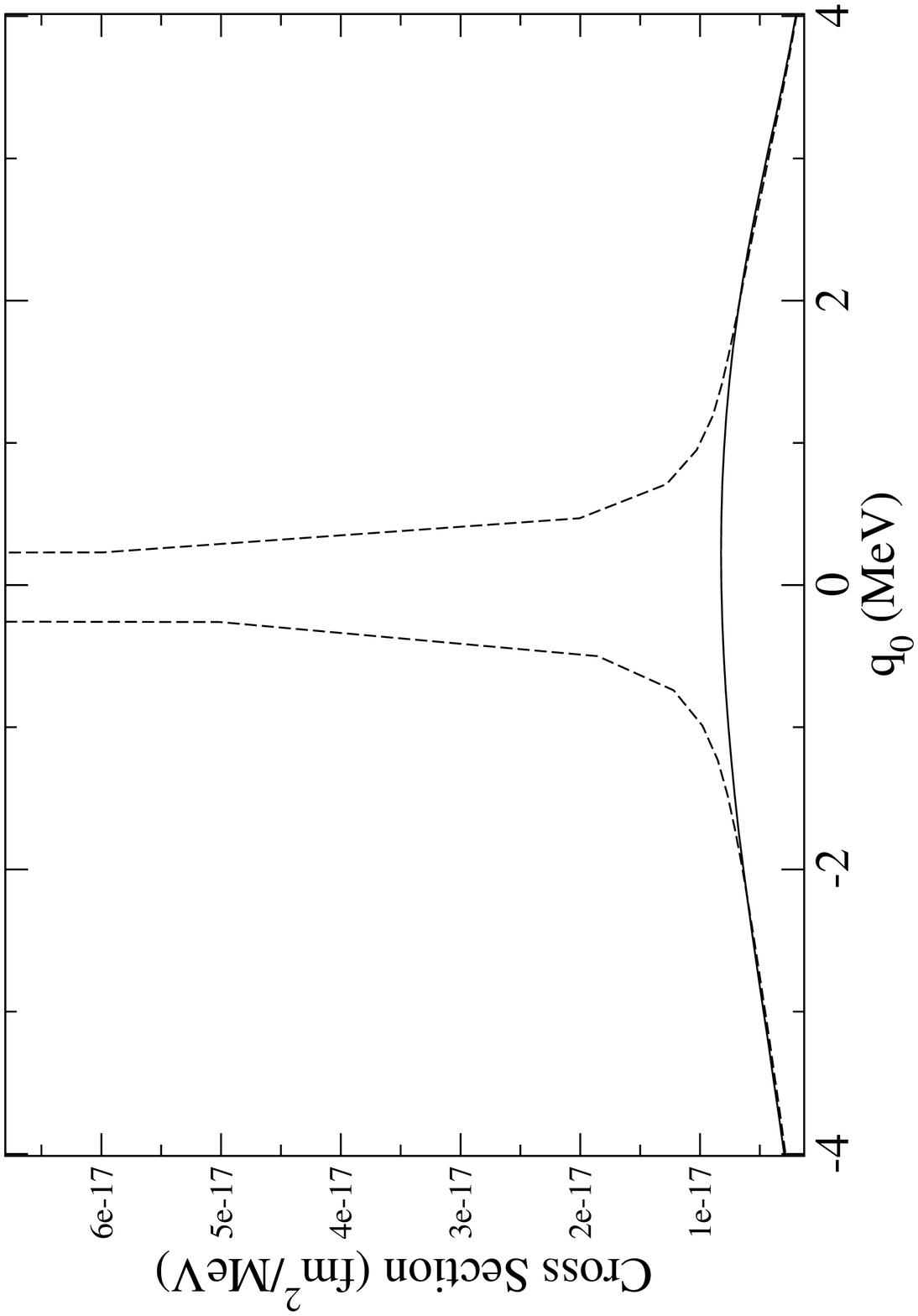}}\hss}}
\caption{Differential cross section $d\sigma/d\Omega dE'$ versus excitation energy $q_0=E_\nu-E_\nu'$ for a baryon density $\rho=0.01 fm^{-3}$, a neutrino energy $E_\nu=20$ MeV, a scattering angle of 90 degrees, an electron fraction $Y_e=0.3$ and a temperature of 5MeV.  The solid curve is in a Hartree approximation while the dashed curve includes RPA correlations.}
\label{fig1}
\end{figure}

\begin{figure}
\vbox to 3.9in{\vss\hbox to 8in{\hss {\includegraphics{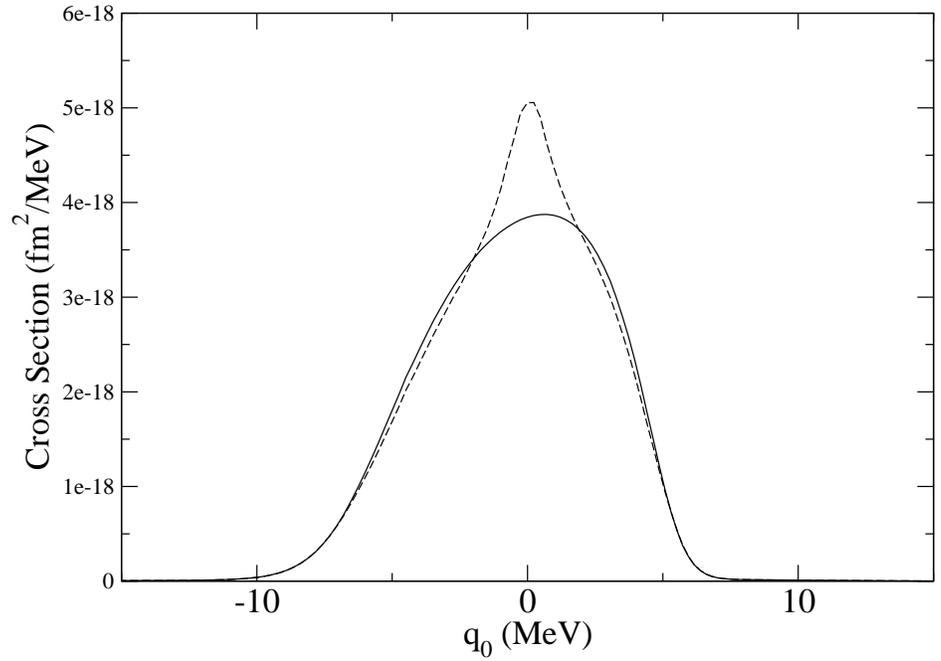}}\hss}}
\caption{As per figure 1 at a baryon density $\rho=0.03 fm^{-3}$.}
\label{fig2}
\end{figure}

\begin{figure}
\vbox to 3.9in{\vss\hbox to 8in{\hss {\includegraphics{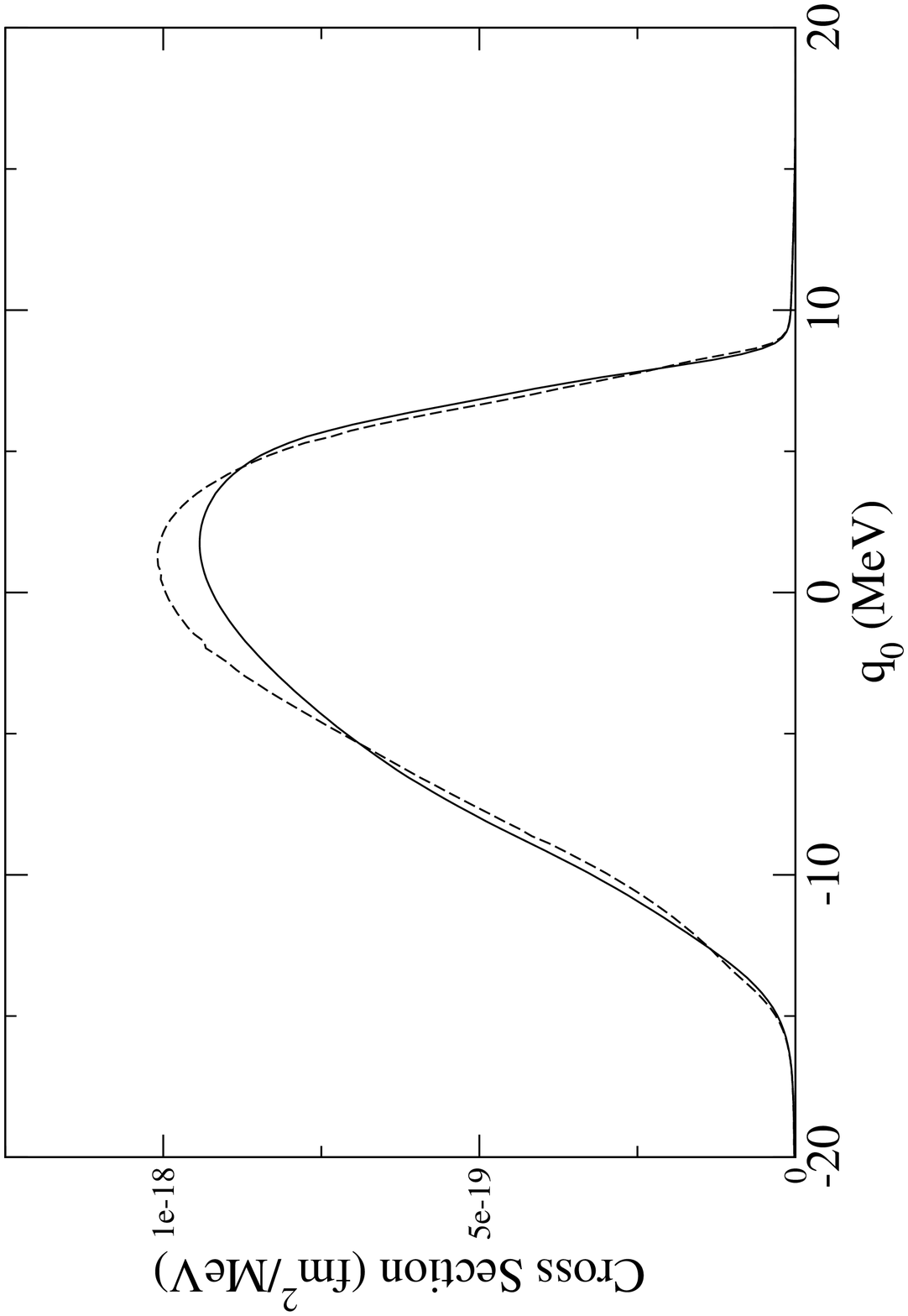}}\hss}}
\caption{As per figure 1 at a baryon density of $\rho=0.1 fm^{-3}$.}
\label{fig3}
\end{figure}

The differential cross section $d\sigma/d\Omega dE'$ for a 20 MeV neutrino scattering at 90 degrees from neutron rich matter is shown in Figures (\ref{fig1},\ref{fig2},\ref{fig3}) at three low densities.  The electron fraction is $Y_e=0.3$ and the temperature is $T=5$ MeV.  At a density of $\rho=0.01$ fm$^{-3}$ the uniform phase is unstable.  This shows up as a large peak in the RPA cross section near $q_0=0$.  This peak decreases as the density increases until it almost disappears by $\rho=0.1$ fm$^{-3}$ where the uniform phase is stable.  

Explicit calculations of opacities for the nonuniform phase would be very useful and are an important area for future work.  We expect the opacity to be greatly enhanced because (1) neutrinos can scatter coherently off of the density changes.  For example Reddy et al find greatly increased opacities for a nonuniform Kaon condensed phase \cite{kaon}.  We note that the neutrons in neutron rich matter carry large weak charges and the neutrino wavelength is comparable to the size of the density fluctuations.  (2) There should be many collective modes at low excitation energies because there may be many different ``pasta" configurations and shapes that all have very similar energies.  The first hint of these modes shows up in RPA calculations as the excitation energy of collective density oscillations drops to zero.  Neutrinos may be able to efficiently excite these modes.  Furthermore, neutral current excitation of these modes may allow mu and tau neutrinos to more efficiently exchange energy with matter and remain in thermal equilibrium to lower densities and temperatures.  

\begin{figure}
\vbox to 3.9in{\vss\hbox to 8in{\hss {\includegraphics{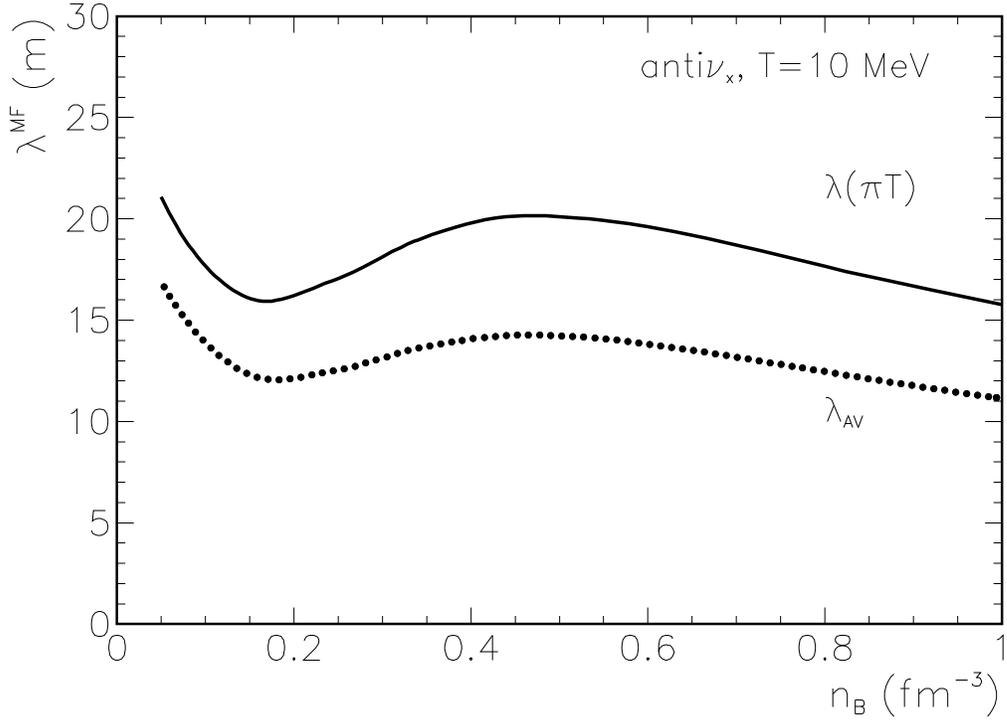}}\hss}}
\caption{Transport mean free path in meters for $\bar\nu_\mu$ or $\bar\nu_\tau$ at a temperature of T=10 MeV as a function of baryon density for neutron rich matter in beta equilibrium.  The solid curve is for a neutrino energy of $E_\nu=\pi T=31.4$ MeV while the dotted curve is the mean free path averaged over neutrino energy.  Both calculations are in a hartree approximation.}
\label{fig4}
\end{figure}

Figure (\ref{fig4}) shows the transport mean free path in a Hartree approximation for $\bar\nu_\mu$ and $\bar\nu_\tau$ versus density.  The energy averaged mean free path Eq. (\ref{eave}) is seen to be somewhat smaller than the mean free path evaluated at an average energy $E=\pi T$.  This could be because Pauli blocking changes the energy dependence of $\lambda(E)$ from $1/E^2$.  Note, in all of the following figures we plot energy averaged quantities.

\begin{figure}
\vbox to 3.8in{\vss\hbox to 8in{\hss {\includegraphics{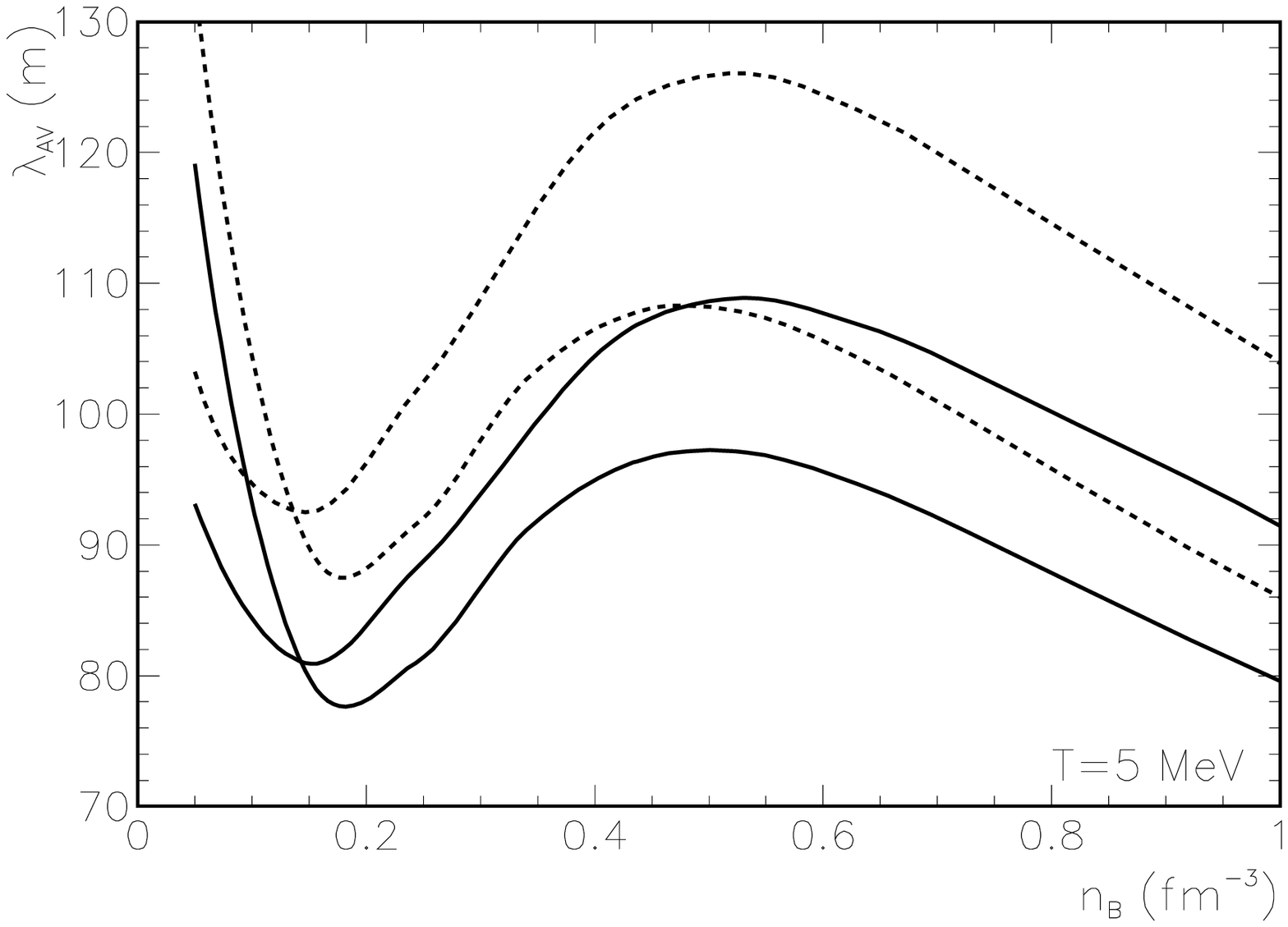}}\hss}}
\caption{Transport mean free path in meters averaged over neutrino energy for neutron rich matter in beta equilibrium at a temperature of $T=5$ MeV.  The solid curves are for $\nu_\mu$ and $\nu_\tau$ while the dotted curves are for $\bar\nu_\mu$ or $\bar\nu_\tau$.  At high density, the upper curves are RPA and the lower curves Hartree calculations.  Note, that the RPA mean free paths are smaller at low density than the Hartree ones. }
\label{fig5}
\end{figure}

Figure (\ref{fig5}) shows the transport mean free path for scattering from neutron rich matter in beta equilibrium at a temperature of 5 MeV.  Note that the mean free path actually increases with density between $n_B\approx 0.2$ and $\approx 0.5$ fm$^{-3}$ because of the rapid decrease of $M^*$ with density.   Weak magnetism causes the mean free path for neutrinos to be systematically smaller than that for antineutrinos.  At high densities, RPA correlations increase the mean free path somewhat.  Our calculations use the simplest RPA interaction that is consistent with the TM1 equation of state.   We do not include residual interactions from $\pi$ exchange or the effects of short range correlations.  These may increase the RPA mean free path further at high densities \cite{relrpa}.  Unfortunately, the parameter $g'$ which was used in \cite{relrpa} to include short range correlations may have some density dependence.  This should be investigated in future work.  Nevertheless, we expect the significant differences between neutrino and antineutrino mean free paths to persist in calculations using more sophisticated interactions.  At lower densities, the RPA mean free path becomes shorter than that in a Hartree approximation because of coupling to density fluctuations related to the transition to a nonuniform phase.  This was previously discussed.

\begin{figure}
\vbox to 3.8in{\vss\hbox to 8in{\hss {\includegraphics{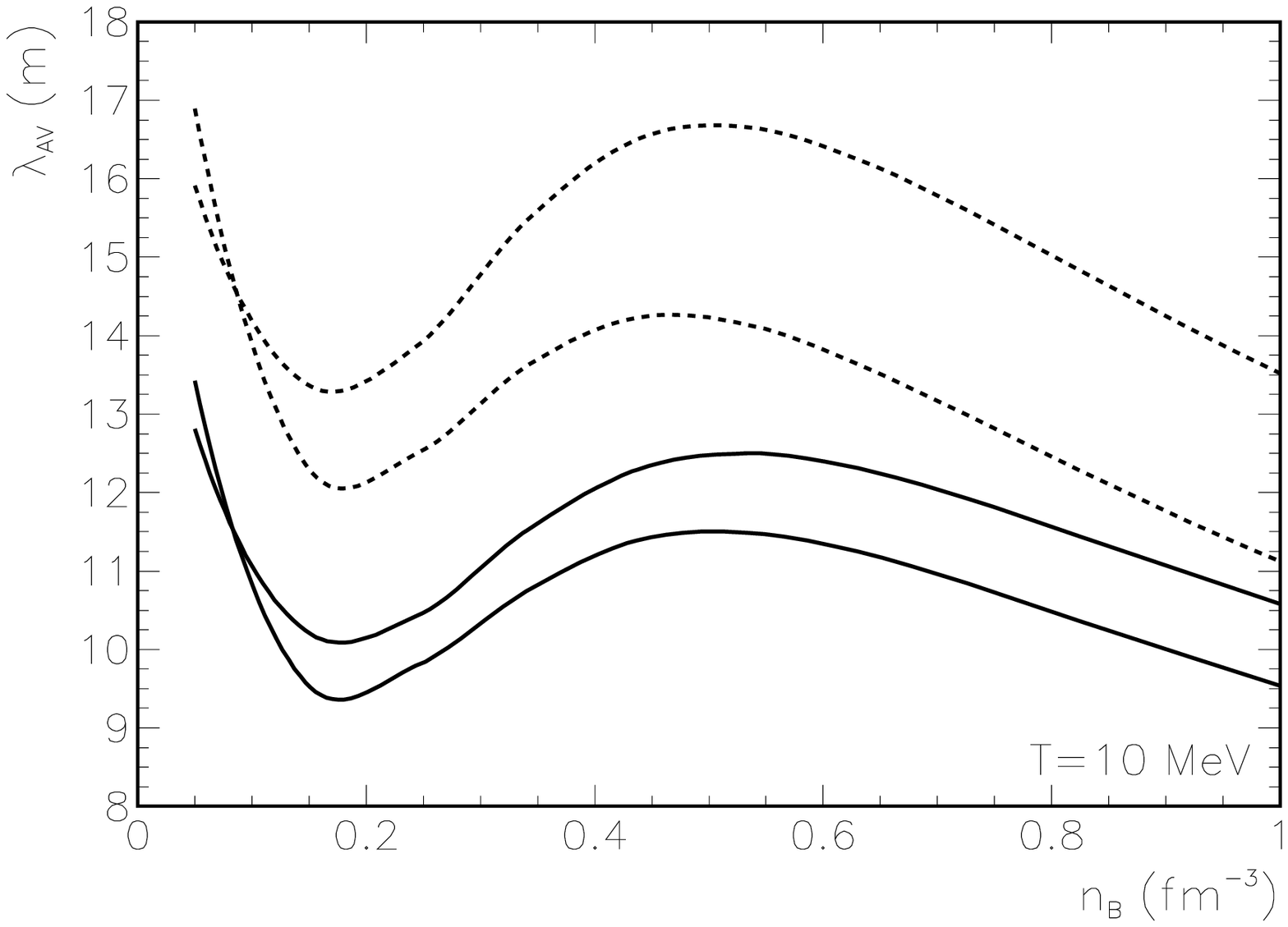}}\hss}}
\caption{Transport mean free path in meters averaged over neutrino energy for neutron rich matter in beta equilibrium at a temperature of $T=10$ MeV.  The solid curves are for $\nu_\mu$ and $\nu_\tau$ while the dotted curves are for $\bar\nu_\mu$ or $\bar\nu_\tau$.  At high density, the upper curves are RPA and the lower curves Hartree calculations.  Note, that the RPA mean free paths are smaller at low density than the Hartree ones. }
\label{fig6}
\end{figure}

\begin{figure}
\vbox to 3.9in{\vss\hbox to 8in{\hss {\includegraphics{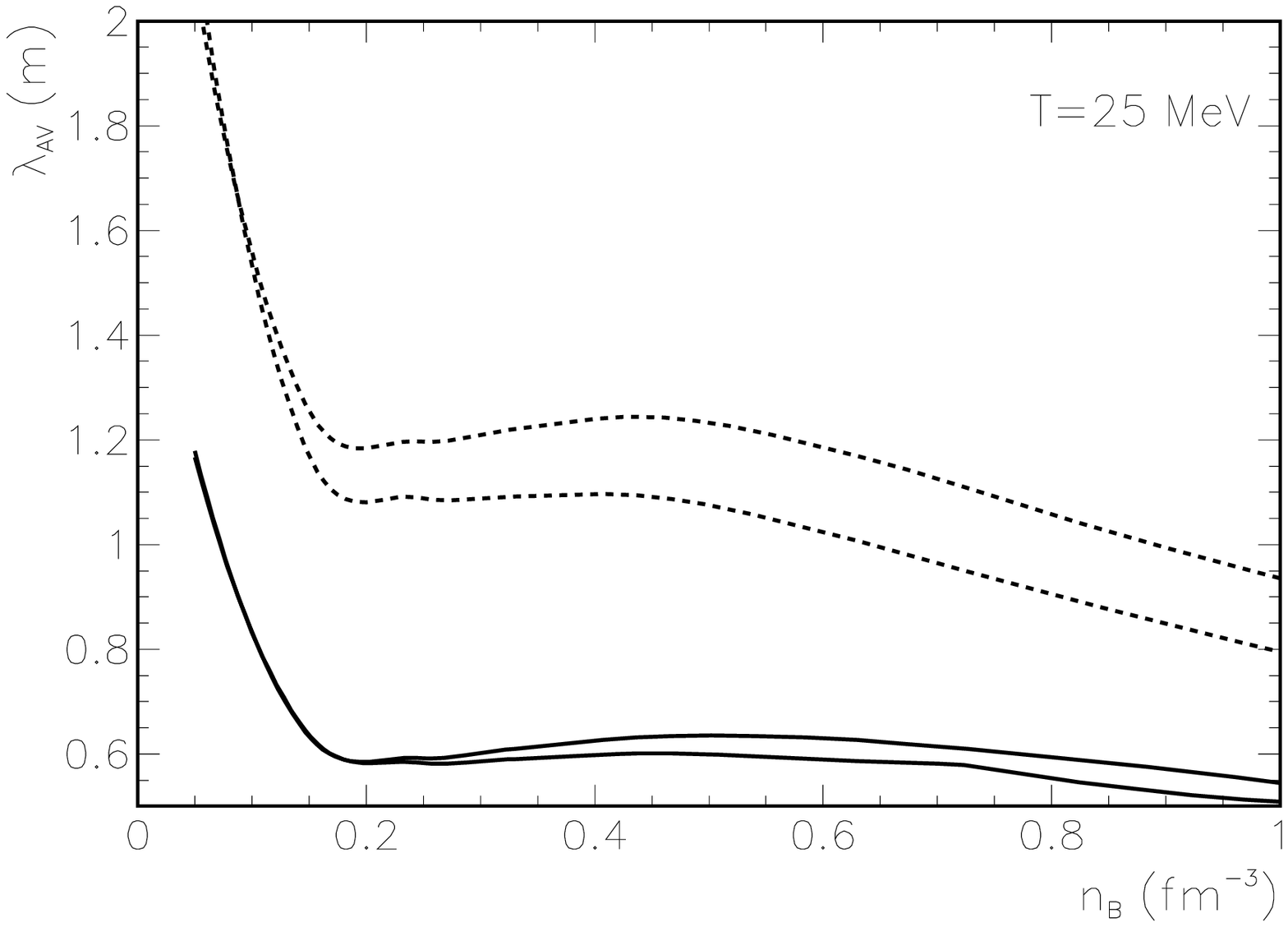}}\hss}}
\caption{Transport mean free path in meters averaged over neutrino energy for neutron rich matter in beta equilibrium at a temperature of $T=25$ MeV.  The solid curves are for $\nu_\mu$ and $\nu_\tau$ while the dotted curves are for $\bar\nu_\mu$ or $\bar\nu_\tau$.  At high density, the upper curves are RPA and the lower curves Hartree calculations.}
\label{fig7}
\end{figure}

\begin{figure}
\vbox to 3.9in{\vss\hbox to 8in{\hss {\includegraphics{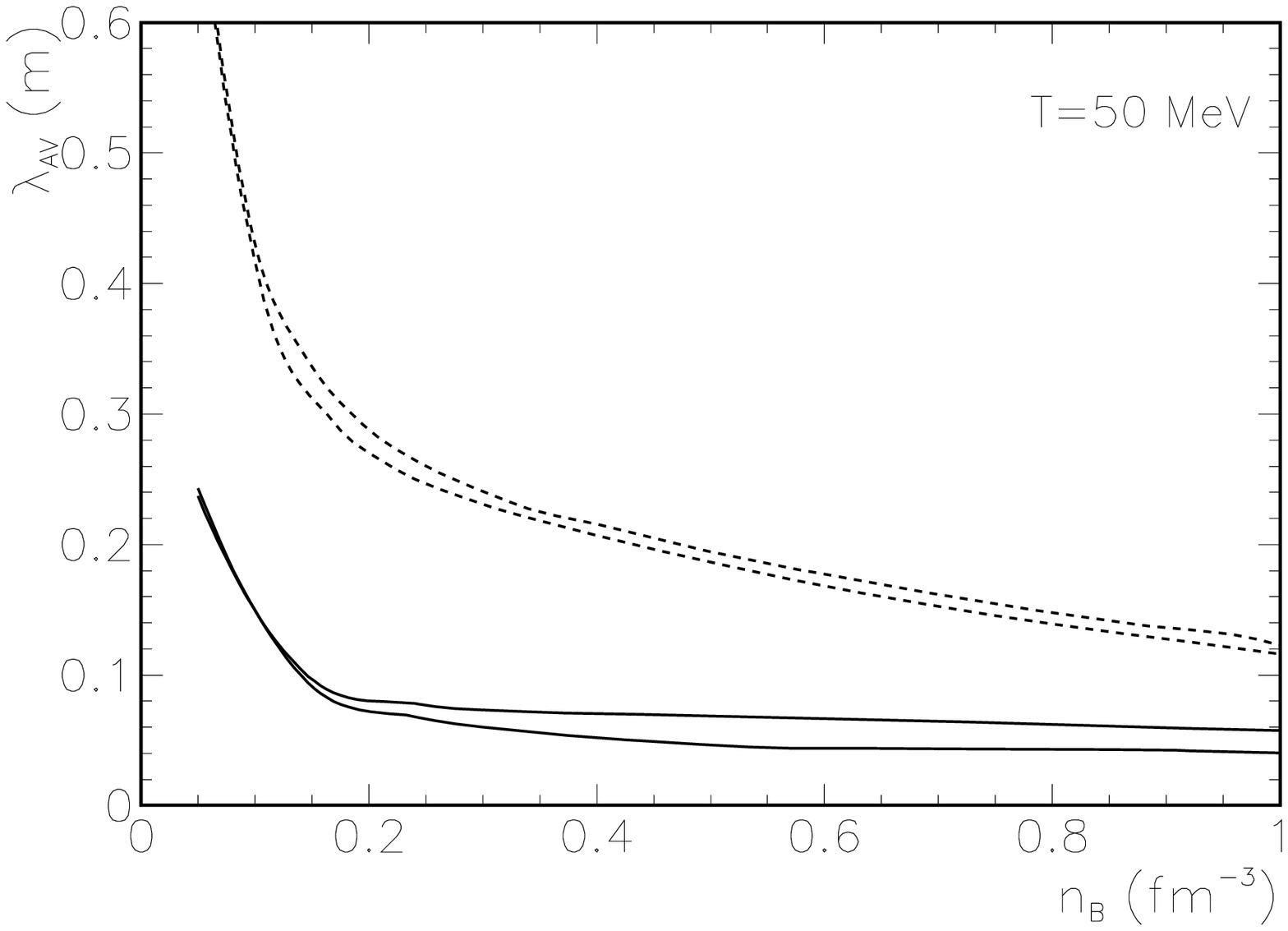}}\hss}}
\caption{Transport mean free path in meters averaged over neutrino energy for neutron rich matter in beta equilibrium at a temperature of $T=50$ MeV.  The solid curves are for $\nu_\mu$ and $\nu_\tau$ while the dotted curves are for $\bar\nu_\mu$ or $\bar\nu_\tau$.  At high density, the upper curves are RPA and the lower curves Hartree calculations.}
\label{fig8}
\end{figure}

Increasing the temperature to 10 MeV, as shown in Figure (\ref{fig6}), increases the average neutrino energy.  Weak magnetism effects grow with energy so the difference between neutrino and antineutrino mean free paths is now larger than at T=5 MeV.  This difference continues to grow as the temperature is increased further to 25 MeV in Figure (\ref{fig7}) and 50 MeV in Figure (\ref{fig8}).  At these higher temperatures there is no longer a significant reduction in the RPA relative to Hartree mean free paths at low densities.  This is because the system is now far from any phase transition.

\begin{figure}
\vbox to 3.9in{\vss\hbox to 8in{\hss {\includegraphics{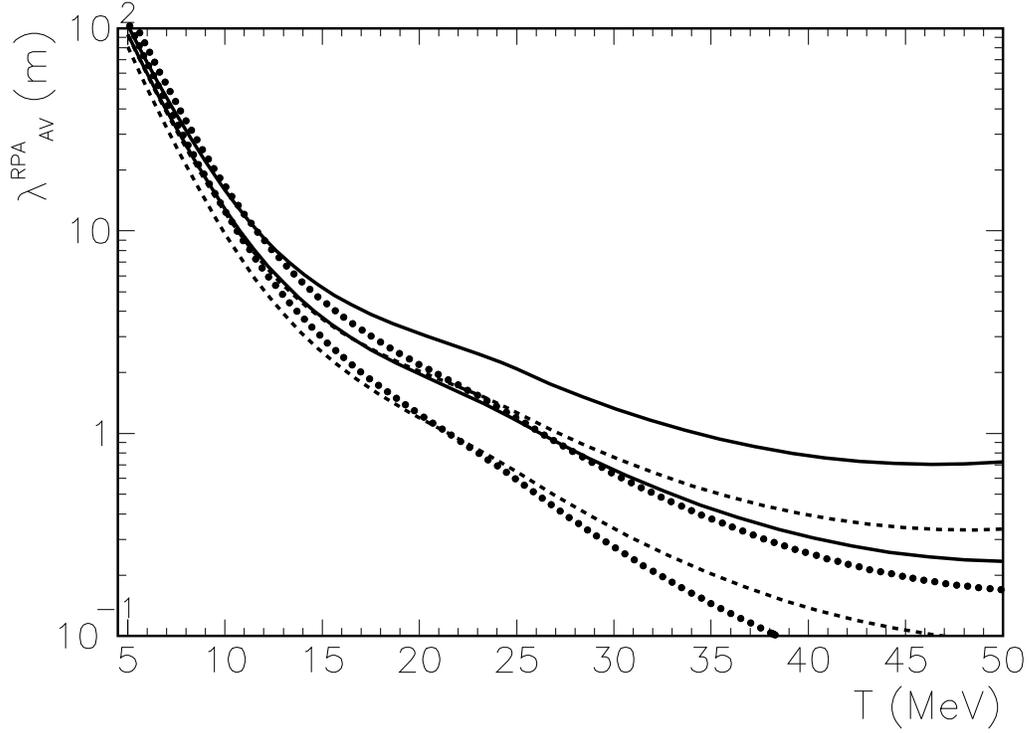}}\hss}}
\caption{Transport mean free path in meters averaged over neutrino energy for neutron rich matter in beta equilibrium versus temperature in an RPA approximation.  The baryon density is 0.05 $fm^{-3}$ for the solid curves, 0.15 $fm^{-3}$ for the dashed and 0.53 $fm^{-3}$ for the dotted curves.  The upper curves are for $\bar\nu_\mu$ and $\bar\nu_\tau$ while the lower curves are for $\nu_\mu$ and $\nu_\tau$.}
\label{fig9}
\end{figure}

\begin{figure}
\vbox to 3.9in{\vss\hbox to 8in{\hss {\includegraphics{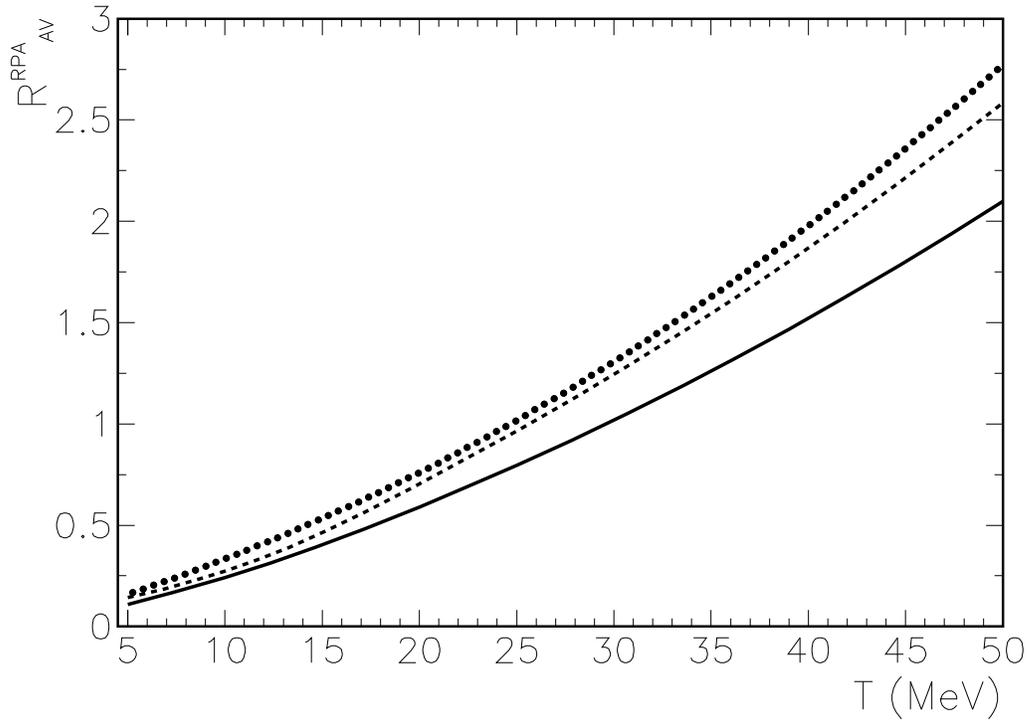}}\hss}}
\caption{Fractional difference between antineutrino and neutrino mean free paths averaged over energy versus temperature in an RPA approximation.   See Eq. (\ref{R}).  The baryon density is 0.05 $fm^{-3}$ for the solid, 0.15 $fm^{-3}$ for the dashed and 0.53 $fm^{-3}$ for the dotted curve.}
\label{fig10}
\end{figure}

Figure (\ref{fig9}) shows the temperature dependence of the mean free paths and Figure (\ref{fig10}) shows the fractional difference $R$ between antineutrino and neutrino mean free paths versus temperature, see Eq. (\ref{R}).  This difference is large and grows rapidly with temperature.  At high temperatures the antineutrinos have much larger mean free paths.  Our RPA calculation has a weak density dependence.  This suggests that the simple weak magnetism correction factors calculated in Ref. \cite{weakmag} at zero density may be a reasonable first approximation to results including correlations in dense matter.  The large ratio $R$ may lead to $\nu_\mu$ and $\nu_\tau$ spectra that are somewhat colder than spectra for $\bar\nu_\mu$ and $\bar\nu_\tau$.

\section{Conclusions}

Advances in neutrino transport allow realistic neutrino interactions to be incorporated into supernova simulations.  This is important because of the very large energy radiated in neutrinos.  We believe a minimum requirement for realistic neutrino opacities is that: (a) they agree, at low densities, with free space neutrino-nucleon cross sections that include weak magnetism and recoil corrections.  (b) Neutrino opacities should be consistent with the equation of state.  To our knowledge, no present simulations have satisfied these basic requirements.

In this paper we have added weak magnetism corrections to relativistic RPA calculations.  Our results agree at low density with accurate neutrino-nucleon cross sections including the fundamental differences between neutrino and antineutrino interactions.  Furthermore, our opacities are consistent with the relativistic mean field equation of state based on the TM1 interaction.

We find antineutrino mean free paths that are considerably larger than those for neutrinos.  This is especially true at high temperatures where neutrino energies are large.  This difference, between $\nu$ and $\bar\nu$ mean free paths, depends only weakly on density and could produce $\nu_\mu$ and $\nu_\tau$ spectra that are somewhat colder than those of $\bar\nu_\mu$ and $\bar\nu_\tau$.

We find RPA correlations somewhat increase mean free paths at high densities.  However, we do not include any residual interaction from pion exchange or short range correlations.  These could increase the mean free path further.  Therefore our results represent a minimum calculation that is consistent with the equation of state.  Future work studying the role of short range correlations and the density dependence of the effective interaction would be very useful.

At low density, correlations reduce the mean free path because of excitation of low energy collective density oscillations.  These are related to the eventual transition of neutron rich matter to a nonuniform phase.  Our RPA calculations can only describe very small amplitude density fluctuations.  Nevertheless we expect neutrinos will couple very effectively to the large amplitude density fluctuations present in the nonuniform phase.  This could lead to a large decrease in the neutrino mean free path and change the emitted neutrino spectrum. We believe, it is important to explicitly calculate neutrino opacities for the nonuniform phase.

\acknowledgments
M.A.P.G. acknowledges the partial financial support of the FICYT and Indiana University.  This work was supported in part by DOE grant DE-FG02-87ER40365.

\section{Appendix: Polarization insertions with tensor couplings}

In this section we provide relations for polarization insertions with tensor couplings.  These compliment previously published results for polarizations with vector couplings.  See for example \cite{HWT} at finite temperature and \cite{HW,hk} at zero temperature.  We use the axial-axial polarization,
\begin{equation}
\Pi_{aa}^{\mu\nu} = -i \int {d^4p\over (2\pi)^4} {\rm Tr} [ G(p+q) \gamma^\mu \gamma_5 G(p) \gamma^\nu \gamma_5],
\end{equation}
that can be written,
\begin{equation}
\Pi_{aa}^{\mu\nu} = \Pi_{vv}^{\mu\nu} + g^{\mu\nu} \Pi_A,
\end{equation}
which defines $\Pi_A$.  Using this we have,
\begin{equation}
\Pi_{vtT} = -\Pi_{vtL} = {q_\mu^2\over 4 M M^*} \Pi_A\,.
\end{equation}
For tensor-tensor correlations we will only need the imaginary parts.
\begin{equation}
{\rm Im} \Pi_{ttT} = -{q_\mu^2\over 4M^2} {\rm Im} 
\Bigl[ \Pi_{vvT} - ( 1 + {q_\mu^2\over 4{M^*}^2} ) \Pi_A \Bigr]
\end{equation}
\begin{equation}
{\rm Im} \Pi_{ttL} = -{q_\mu^2\over 4M^2} {\rm Im} \Bigl[ \Pi_{vvL} + 
( 1 + {q_\mu^2\over 4{M^*}^2} ) \Pi_A \Bigr]
\end{equation}
The vector-axial polarization,
\begin{equation}
\Pi_{va}^{\mu\nu} =-i \int {d^4p\over (2\pi)^4}  {\rm Tr} [G(p+q) \gamma^\mu G(p) \gamma^\nu\gamma_5 ] 
\end{equation}
can be written,
\begin{equation}
\Pi_{va}^{\mu\nu}= -i \epsilon^{\mu\nu\alpha 0} q_\alpha \Pi_{va}\,.
\end{equation}
The tensor-axial polarization is simply related to $\Pi_{va}$,
\begin{equation}
\Pi_{ta} = {M^*\over M} \Pi_{va} \,.
\end{equation}
This has been used in Eq. (\ref{r3}).  Finally, the mixed scalar-tensor polarization is simply related to the scalar-vector polarization,
\begin{equation}
\Pi_{Mt}^\mu={q_\mu^2\over 4MM^*} \Pi_M^\mu\,.
\end{equation}
This has been incorporated into Eq. (\ref{mixedW}).


\newpage

\end{document}